\documentclass[useAMS, usenatbib,referee]{biom}
\usepackage{mathtools}
\usepackage{amsfonts}
\let\theoremstyle\undefined

\usepackage{amsthm}
\usepackage[colorlinks,citecolor=blue,urlcolor=blue]{hyperref}

\usepackage{booktabs}
\usepackage{diagbox}
\usepackage{multirow}
\usepackage{algorithm}
\usepackage{algpseudocode}
\usepackage{xpatch}

\def\>={\geqslant}
\def\<={\leqslant}
\newcommand{\Prob}{\mathbb{P}\kern\nulldelimiterspace}
\newcommand{\mat}[1]{\mathbf{#1}}
\renewcommand{\vec}[1]{\bmath #1}
\newcommand{\T}{{\mkern-1.5mu\mathsf{T}}}
\newcommand{\KER}{\hat{K}_{\hspace{-1pt}E\hspace{-1pt}R}}
\newcommand{\distApprox}{\mathrel{\dot\sim}}

\newcommand{\aln}[3][c]{
	\mathmakebox[\widthof{$\mathsurround=0pt #2$}][#1]{#3}
}

\makeatletter
\newcommand\myvec[1]{%
	\@tempcnta=0
	\@for\@ii:=#1\do{%
		\@insertbreakingcomma
		\@ii
	}%
}
\def\@insertbreakingcomma{%
	\ifnum \@tempcnta = 0 \else, \linebreak[1] \fi
	\advance\@tempcnta\@ne
}
\makeatother

\newtheoremstyle{colon}{}{}{\itshape}{}{\scshape}{\ \bfseries A\scshape:}{ }{}
\theoremstyle{colon}
\newtheorem*{asymRegime}{Asymptotic Regime}

\allowdisplaybreaks

\hypersetup{
	pdflinkmargin=0pt, 
	colorlinks=true,
} 
\usepackage{etoolbox}

\makeatletter

\DeclareRobustCommand\citepos
{\begingroup\def\NAT@nmfmt##1{{\NAT@up##1's}}%
	\NAT@swafalse\let\NAT@ctype\z@\NAT@partrue
	\@ifstar{\NAT@fulltrue\NAT@citetp}{\NAT@fullfalse\NAT@citetp}}

\pretocmd{\NAT@citex}{%
	\let\NAT@hyper@\NAT@hyper@citex
	\def\NAT@postnote{#2}%
	\setcounter{NAT@total@cites}{0}%
	\setcounter{NAT@count@cites}{0}%
	\forcsvlist{\stepcounter{NAT@total@cites}\@gobble}{#3}}{}{}
\newcounter{NAT@total@cites}
\newcounter{NAT@count@cites}
\def\NAT@postnote{}

\def\NAT@hyper@citex#1{%
	\stepcounter{NAT@count@cites}%
	\hyper@natlinkstart{\@citeb\@extra@b@citeb}#1%
	\ifnumequal{\value{NAT@count@cites}}{\value{NAT@total@cites}}
	{\if*\NAT@postnote*\else\NAT@cmt\NAT@postnote\global\def\NAT@postnote{}\fi}{}%
	\ifNAT@swa\else\if\relax\NAT@date\relax
	\else\NAT@@close\global\let\NAT@nm\@empty\fi\fi
	\hyper@natlinkend}
\renewcommand\hyper@natlinkbreak[2]{\@strut #1}

\patchcmd{\NAT@citex}
{\ifNAT@swa\else\if*#2*\else\NAT@cmt#2\fi
	\if\relax\NAT@date\relax\else\NAT@@close\fi\fi}{}{}{}
\patchcmd{\NAT@citex}
{\if\relax\NAT@date\relax\NAT@def@citea\else\NAT@def@citea@close\fi}
{\if\relax\NAT@date\relax\NAT@def@citea\else\NAT@def@citea@space\fi}{}{}
\patchcmd{\NAT@cite}{\if*#3*}{\if*\NAT@postnote*}{}{}

\newbox\@strutbox
\setbox\@strutbox\hbox{%
	\vrule\@height.65\baselineskip%
	\@depth.25\baselineskip%
	\@width\z@}%
\def\@strut{\relax\ifmmode\copy\@strutbox\else\unhcopy\@strutbox\fi}

\newcommand{\refmail}[1]{\@strut \href{mailto:#1}{\nolinkurl{#1}}\@strut}
\newcommand{\refurl}[2][]{%
	\@strut%
	\ifstrempty{#1}{%
		\url{#2}%
	}{%
		\href{#2}{#1}%
	}%
	\@strut%
}

\newrobustcmd\ref@wrapper[5][]{%
	\@strut 
	\ifstrempty{#2}{
		\hyperref[#3]{#1#4}%
	}{
		\hyperref[#2]{#1\hypersetup{pdfborder={0 0 0}}#5}%
	}%
	\@strut%
}
\newcommand{\reffloat}[3][]{\ref@wrapper[#3]{#1}{#2}{\ref*{#2}}{\ref{#1}-\ref{#2}}}

\makeatother

\algrenewcommand\alglinenumber[1]{\footnotesize #1\phantom{:}} 
\renewcommand{\thealgorithm}{} 

\makeatletter
\xpatchcmd{\algorithmic}{\itemsep\z@}{\itemsep=0.5ex plus2pt}{}{}

\newenvironment{breakablealgorithm}
{
	\begin{center}
		\refstepcounter{algorithm}
		\hrule height.8pt depth0pt \kern3pt
		\renewcommand{\caption}[2][\relax]{
			{\raggedright\textbf{\ALG@name~\thealgorithm} ##2\par}%
			\ifx\relax##1\relax 
			\addcontentsline{loa}{algorithm}{\protect\numberline{\thealgorithm}##2}%
			\else 
			\addcontentsline{loa}{algorithm}{\protect\numberline{\thealgorithm}##1}%
			\fi
			\kern2pt\hrule\kern2pt
		}
	}{
		\kern2pt\hrule\relax
	\end{center}
}
\makeatother

\title[Eigenvalue Ratio Approach to Inferring Population Structure from Sequencing Data]
{An Eigenvalue Ratio Approach to Inferring Population Structure from Whole Genome Sequencing Data}

\author{
Yuyang Xu$^{1,*}$\email{\refmail{xuyy@connect.hku.hk}}, 
Zhonghua Liu$^{1,**}$\email{\refmail{zhhliu@hku.hk}}, and 
Jianfeng Yao$^{2,***}$\email{\refmail{jeffyao@cuhk.edu.cn}} \\
$^{1}$Department of Statistics and Actuarial Science, The University of Hong Kong, Hong Kong SAR, China \\
$^{2}$School of Data Science, The Chinese University of Hong Kong (Shenzhen), Shenzhen, China
}

\begin{document}


\date{{\it Received July} 2021. {\it Revised April} 2022. {\it
Accepted April} 2022.}



\pagerange{\pageref{firstpage}--\pageref{lastpage}} 
\volume{64}
\pubyear{2022}
\artmonth{April}


\doi{10.1111/j.1541-0420.2005.00454.x}


\label{firstpage}


\begin{abstract}
Inference of population structure from genetic data plays an important role in population and medical genetics studies. With the advancement and decreasing cost of sequencing technology, the increasingly available whole genome sequencing data provide much richer information about the underlying population structure. The traditional method \citep*{patterson2006population} originally developed for array-based genotype data for computing and selecting top principal components that capture population structure may not perform well on sequencing data for two reasons. First, the number of genetic variants $p$ is much larger than the sample size $n$ in sequencing data such that the sample-to-marker ratio $n/p$ is nearly zero, violating the assumption of the Tracy--Widom test used in their method. Second, their method might not be able to handle the linkage disequilibrium well in sequencing data. To resolve those two practical issues, we propose a new method called ERStruct to determine the number of top informative principal components based on sequencing data. More specifically, we propose to use the ratio of consecutive eigenvalues as a more robust test statistic, and then we approximate its null distribution  using modern random matrix theory. Both simulation studies and applications to two public data sets from the HapMap 3 and the 1000 Genomes Projects demonstrate the empirical performance of our ERStruct method. 
\end{abstract}

%



\maketitle


%

\section{Introduction} \label{sec:introduction}
Inference of population structure is a fundamental problem in population genetics and also plays a critical role in genetic association studies using whole genome sequencing data.  For example, in population genetic studies, \citet{wu2019large} and \citet{cao2020chinamap} performed principal component analysis (PCA) using  whole genome sequencing data for the discovery of population genetic diversity in China and Singapore, respectively. In genetic association studies, the presence of population stratification may lead to spurious  association estimates \citep{price2006principal, mathieson2012differential, wang2014ancestry}. It is thus critical to control for the underlying population structure to avoid spurious associations when mapping the genetic basis for complex traits and human diseases

PCA-based methods have been popularized to capture the population structure from array-based genotype data \citep*{menozzi1978synthetic, patterson2006population, reich2008principal}. These methods compute and select top principal components (PCs) that can sufficiently capture population structure \citep{patterson2006population}. Then the selected PCs can be  further used to correct for population stratification bias in genetic association studies, for example, using the popular EIGENSTRAT method \citep{price2006principal}.

A key question when applying PCA to genetic data is how to determine the number of PCs that can sufficiently capture the underlying unknown population structure. \citet{patterson2006population} modified the original Tracy--Widom (TW) test \citep{tracy1994level, johnstone2001distribution} and proposed the so-called sequential Tracy--Widom test using the effective (reduced) number of markers as a plug-in estimate to replace the original number of markers in the array-based genotype data set. The effective number of markers is used to estimate the number of the underlying uncorrelated markers so that the number of markers used for the Tracy--Widom test can be effectively reduced. With the use of effective number of markers, \citet{patterson2006population} tried to alleviate the possible violation of the assumption in the Tracy--Widom test that the sample size and the number of genetic markers should be comparably large. We will refer to this method as the PCA-TW test throughout this paper. 

However, after being applied in various empirical studies for several years, it has been found that the PCA-TW test might not perform well for capturing the true population structure in sequencing data  \citep*{zhang2013association, zhang2013sequencing, zhou2018eigenvalue}. This is because the traditional PCA-TW test was originally developed for array-based genotype data sets that typically contain a moderate-to-high number of genetic markers, while the number of genetic markers is much larger in sequencing data. As large-scale sequencing data sets become increasingly available \citep{10002015global, bycroft2018uk, wu2019large, cao2020chinamap}, it is thus pressing to develop a new method that can resolve the following two   practically important issues:

\begin{enumerate}
	\item \emph{Ultra-dimensionality\/} (or ultra-high-dimensionality), which refers to the scenario in which the sample-to-marker ratio $n/p$ is nearly zero. In the random matrix theory literature, this is essentially a different regime from the one used in the PCA-TW test which assumes that $n$ and $p$ are comparably large \citep{johnstone2001distribution}. A genotype sequencing data set typically includes millions of markers, which makes the ratio $n/p$ goes to an order of $10^{-4}$ or even smaller. In the ultra-dimensional settings, the ad hoc approach of using the effective number of markers might not perform well for sequencing data.
	
	\item \emph{Linkage disequilibrium\/} (LD). Genetic markers in a sequencing data set may have very high correlations ($0.6\,$--$\,0.9$). This issue becomes even worse when markers are physically close to each other on the chromosome. The theoretical validity of the PCA-TW test requires the independence assumption among the genetic markers. Hence, the presence of LD may seriously distort the null distribution of the test statistic and thus leads to biased inference. As \citet{patterson2006population} pointed out in their paper, there are several issues when applying their method  on data sets with large admixture-LD. To correct for the presence of LD, \citet{patterson2006population} recommended a modification of their PCA-TW test method using backward regression. However, this correction is computationally intensive, especially when a wide range of genetic markers are in LD with each other. Another method is LD pruning \citep{purcell2007plink, bycroft2018uk, zhou2018eigenvalue,cao2020chinamap}, which removes genetic markers based on high levels of pairwise LD. This LD pruning method apparently will result in a loss of information about population structures as it might remove ancestry informative markers.
\end{enumerate}

So far, several extensions of the PCA-TW test have been proposed. \citet{shriner2012improved} proposed an alternative plug-in estimate of the effective number of markers for the PCA-TW test. However, the reason why choosing such a plug-in estimate has not been theoretically justified. \citet{zhou2018eigenvalue} proposed two methods to improve the PCA-TW test. The first one is the model-based method that tries to reduce the influence of the LD by correcting for the local correlation structure with an alternative eigenvalue limiting distribution. However, the simple discrete distribution of the population truth in the alternative model is chosen without theoretical justification. Their second method is to use block permutation to find out an appropriate null eigenvalue distribution. But such an approach is computationally intensive and might be computationally expensive for large-scale sequencing data.

The PCA-TW test and its extensions by \citet{shriner2012improved} and \citet{zhou2018eigenvalue} all share one common key idea, that is, the sample covariance matrix can be viewed as a \emph{finite rank perturbation\/} of the sample noise covariance matrix. The theory of finite-rank perturbation of a large random matrix originates from the seminal spiked population model introduced by \citet{johnstone2001distribution}. After that, there are subsequent important developments \citep*{paul2007asymptotics, baik2006eigenvalues, baik2005phase, bai2008central, benaych2011eigenvalues, benaych2011fluctuations}. The theory essentially states that the non-zero ordered eigenvalues of the sample covariance matrix $\mat{S}_n$ can be separated into two parts: (1) the largest few ones are called \emph{spikes\/}, whose number is the same as the number of top informative PCs minus one; (2) the remaining ones are called \emph{bulk\/}, which asymptotically form a dense distribution well-separated from the spikes. Therefore, the null distribution of the top bulk eigenvalues can be used to detect spikes and to estimate the number of top informative PCs. 

Unlike the PCA-TW test and its aforementioned extensions, several researchers proposed to use the ratio (or more generally, ratio-wise functions) of the consecutive eigenvalues ($\ell_1{\>=} \ell_2{\>=} {\cdot}{\cdot}{\cdot}{\>=} \ell_n >0$) as a more robust test statistic to detect spikes. An estimator using the ratio of eigenvalue differences $(\ell_i - \ell_{i+1}) / (\ell_{i+1} - \ell_{i+2})$ as a test statistic was first proposed by \citet{onatski2009testing} to estimate the number of factors. Later, \citet{lam2012factor} and \citet{ahn2013approximate} proposed an eigenvalue ratio (ER) based estimator ${\arg}{\max}_{i\<=K_{\mathrm{max}}} \ell_i/\ell_{i+1}$ given a pre-determined maximum possible number of factors $K_{\mathrm{max}}$. Another type of ER-based estimator $\min\{ i\>=1 \text{ s.t. } \ell_i/\ell_{i+1} > 1{-}d_T \}$ was proposed by \citet*{li2017identifying}, where $d_T$ is chosen within $(0,1)$. These previous results all show substantial improvements for spike detection. However, previous works on ER-based estimators focus mainly on moderate-to-high dimensional data sets with mild correlations among features. Particularly, these works still require that the sample size $n$ and the number of features $p$ to be of comparable magnitude. For example, the return of stocks data used in \citet{li2017identifying} has a sample-to-feature ratio of $16.89$. Thus, those methods are not applicable to modern ultra-dimensional sequencing data sets where $p\gg n$, together with complicated LD structures among genetic markers.	

In this paper, we propose a novel ER-based estimator to infer latent population structure (\emph{ERStruct\/}) from ultra-dimensional sequencing data in the framework of analysis of variance (ANOVA) model by leveraging the fact that different latent sub-populations have different minor allele frequencies (MAF). Although, our ER-based estimator is inspired by \citet{li2017identifying}, however our method makes two new methodological contributions. First, by leveraging the recent theoretical results from random matrix theory \citep{benaych2011eigenvalues, benaych2011fluctuations, wang2014limiting}, we find a new way to approximate the distribution of the eigenvalues of the sample covariance matrix under ultra-dimensionality regime by the distribution of the eigenvalues of a high-dimensional Gaussian orthogonal ensemble (GOE) matrix. Then, we use the known random matrix theory under high-dimensional regime and develop an adaptive approximation to the true null distribution, which also greatly reduces the computational burden. Second, we further resolve the LD problem in the sequencing data sets by proposing new estimates of the parameters in the approximation theory developed by \citet{wang2014limiting} and obtain the LD-adjusted null distribution under the ultra-dimensional regime.
We conduct simulation studies to compare our ERStruct method with the traditional PCA-TW method. Moreover, we apply our ERStruct method to the HapMap 3 data set \citep{international2010integrating} and the 1000 Genomes Project sequencing data set \citep{10002015global}. Our proposed ERStruct method was shown to be   accurate, robust and  also computationally efficient.

The rest of this paper is organized as follows.  In Section~\reffloat{sec:method}{}, we introduce the proposed ERStruct method and its computational algorithm. In Section~\reffloat{sec:simulation_studies}{}, we perform simulation studies to compare our ERStruct method with the traditional PCA-TW test. In Section~\reffloat{sec:real_data_analysis}{}, we apply our ERStruct method to two real data sets and compare its performance with the PCA-TW test. This paper ends with discussions in Section~\reffloat{sec:discussion}{}.

\section{Method} \label{sec:method}
Suppose that one is interested in estimating the number of the top informative PCs (i.e., latent sub-populations) that capture population structures based on a raw $n$-by-$p$ genotype matrix $\mat{C}$ which contains $p$ genetic markers from $n$ individuals. Each entry $\mat{C}(i,j) \in \{0,1,2\}$ represents the raw count of the minor alleles for the genetic marker $j$ on the individual $i$. Assume that there are $K$ (latent) sub-populations and the $k$th sub-population is of size $n_k$, where ${\sum_{k=1}^K} n_k = n$ and $1\<= k \<= K$. To refer to a specific individual within the $k$th sub-population, the index of individual $i$ is rewritten as follows:
\[
i = (k,l),\qquad k=1,\dotsc,K \quad \text{and} \quad l=1,\dotsc,n_k,
\]
where $l$ indexed the individuals within the sub-population. Using this notation, we can rewrite the raw count matrix $\mat{C}$ as
\[
\mat{C} = ( \underbrace{\vec{c}_{1,1}^\T{,...},\vec{c}_{1,n_1}^\T}_{\text{1st sub-pop}},\ \underbrace{\vec{c}_{2,1}^\T{,...},\vec{c}_{2,n_2}^\T}_{\text{2nd sub-pop}},\ \ldots\ldots,\ \underbrace{\vec{c}_{K,1}^\T{,...},\vec{c}_{K,n_K}^\T}_{K\text{th sub-pop}} )^\T,
\]
where $\vec{c}_{k,l} \equiv \mat{C}(i,\cdot)$ is a $p$-dimensional vector containing the values of the genetic markers for the $i$th individual. We consider the following asymptotic regime throughout this paper. 

\begin{asymRegime}
	\[\begin{dcases}
	\text{sample size } n\rightarrow\infty, \\
	\text{sample-to-marker ratio } n/p \rightarrow 0.
	\end{dcases}
	\label{regime}
	\]
\end{asymRegime}

This asymptotic regime is reasonable in whole genome sequencing data, where the number of markers $p$ is much larger than the sample size $n$.

\subsection{Modeling Framework}
Our model builds on the key observation that individuals from different sub-populations have different minor allele frequencies (MAF) and individuals from the same sub-population have the same MAF. This observation motivates us to model the raw minor allele count data matrix using the following analysis of variance (ANOVA) model, 
\begin{equation}
\vec{c}_{k,l} = \vec{\mu}_k + \vec{\varepsilon}_{k,l},\label{eq:ANOVA}
\end{equation}
where the $p$-dimensional vectors $\vec{\mu}_1,\dotsc,\vec{\mu}_K$ are the sub-population-specific mean counts of minor alleles across the $K$ sub-populations, and the vectors $\vec{\varepsilon}_{1,1},\dotsc,\vec{\varepsilon}_{K,n_K}$ are independent and identically distributed noise vectors with mean zeros and covariance $\mat{\Sigma}$. We emphasize here that even though we use this ANOVA model for the raw minor allele count matrix, however, our model differs from the classical ANOVA model in three ways. First, we do not know the total number of sub-populations $K$ a priori and which individual belongs to which sub-population. Second, the sample to-marker-ratio $n/p$ is nearly zero. Third, our model allows for the presence of different LD patterns. Those three salient features of our ANOVA model thus require modern random matrix theory to understand the sources of the variation in the sequencing data. 

Following \citet{patterson2006population}, we normalize the raw count matrix $\mat{C}$ such that each column (genetic marker) has mean zero and unit variance. The estimates of the $k$th sub-population-specific mean vector $\hat{\vec\mu}_k$ and the estimate of the overall mean vector $\hat{\vec\mu}$ are  given by 
\begin{align}
\hat{\vec\mu}_k &= \frac1{n_k} \sum_{l=1}^{n_k}\vec{c}_{k,l}, \label{eq:mu_hat} \\
\hat{\vec\mu} &= \frac1n \sum_{k=1}^{K} \sum_{l=1}^{n_k} \vec{c}_{k,l} = (\hat{\mu}_1{,...},\hat{\mu}_p)^\T. \notag
\end{align}

We also need the following diagonal matrix with diagonal elements equal to the inverse of the standard deviations of the $p$ genetic markers
\begin{gather}
\hat{\mat{D}} = \mathrm{diag}\Bigl(1 \Big/ {\sqrt{\hat{\mu}_j(1{-}\hat{\mu}_j/2)}} \Bigr), \quad 1\<=j\<=p. \notag \\
\intertext{Then the normalized genotype data matrix is given by}
\mat{M} = \bigl( \hat{\mat{D}} {\cdot} (\vec{c}_{1,1}-\hat{\vec\mu}), \dotsc,  \hat{\mat{D}} {\cdot} (\vec{c}_{K,n_K}-\hat{\vec\mu}) \bigr)^\T.
\label{eq:normalization}
\end{gather}

Define the sample covariance matrix of the normalized data as $\mat{S}_n = \mat{M}\mat{M}^\T /n$. Then, we have the standard ANOVA decomposition $\mat{S}_n \equiv \mat{S}_\mathrm{B} + \mat{S}_\mathrm{W}$,
where the between-group variation $\mat{S}_\mathrm{B}$ and the within-group variation $\mat{S}_\mathrm{W}$ are given respectively by 
\begin{align}
	\mat{S}_\mathrm{B} &= \sum_{k=1}^K \frac{n_k}{n} \Bigl\{ \hat{\mat{D}} (\hat{\vec\mu} - \hat{\vec\mu}_k) \Bigr\}^{\!\T} \Bigl\{ \hat{\mat{D}} (\hat{\vec\mu} - \hat{\vec\mu}_k) \Bigr\}, \notag\\
	\mat{S}_\mathrm{W} &= \frac1n \sum_{k=1}^K\sum_{l=1}^{n_k} \Bigl\{ \hat{\mat{D}} (\vec\mu_k - \hat{\vec\mu}_k + \vec{\varepsilon}_{k,l}) \Bigr\}^{\!\T} \Bigl\{ \hat{\mat{D}} (\vec\mu_k - \hat{\vec\mu}_k + \vec{\varepsilon}_{k,l}) \Bigr\}. \label{eq:B_W}
\end{align}
In the next section, we will perform spectral analysis for $\mat{S}_n$ using modern random matrix theory.

\subsection{The Spikes and the Bulk}
\begin{figure}
	\centering
	\includegraphics[width=2.7in]{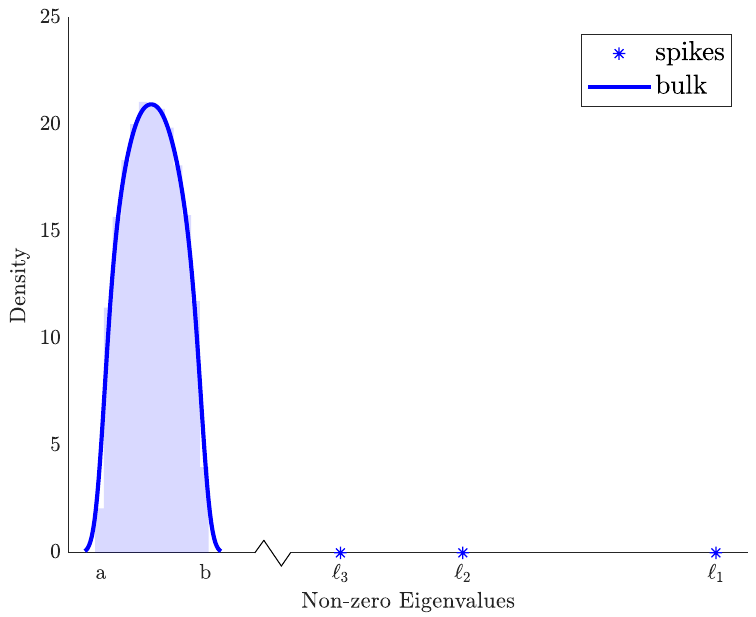}
	\caption{A typical distribution of non-zero sample eigenvalues of sample covariance matrix $\mat{S}_n$ (or equivalently $\mat{S}_p$) computed from an ultra-dimensional data generated based on the uncorrelated simulation setting discussed in Section~\reffloat{sec:simulation_studies}{} (except that for a clearer view, the true number of sub-populations is chosen as $K=4$ and the size of all the sub-populations are $n_k = \{550, 600, 650, 700\}$). The broken line in the middle of the x-axis indicates a big gap between the spikes and the bulk.}
	\label{fig:eg_bulk_spikes}
\end{figure}

It can be seen from Equation~\eqref{eq:B_W}{} that the within-group covariance matrix $\mat{S}_\mathrm{W}$ is essentially the noise covariance matrix. The between-group covariance matrix $\mat{S}_\mathrm{B}$ is of rank $K{-}1$ because it is the sum of $K$ rank one matrices and those $K$ between-group vectors $\bigl\{ \hat{\mat{D}} (\hat{\vec\mu}_k {-} \hat{\vec\mu}) \bigr\}_K$ satisfy the following linear constraint:
\[
\sum_{k=1}^K \frac{n_k}{n}\, \hat{\mat{D}} (\hat{\vec\mu}_k - \hat{\vec\mu}) = \hat{\mat{D}} \cdot \vec{0} = \vec{0}.
\]
Hence the matrix $\mat{S}_n$ can be viewed as a $K{-}1$ rank perturbation of the sample noise covariance matrix $\mat{S}_\mathrm{W}$. According to the finite-rank perturbation theory \citep{benaych2011eigenvalues, benaych2011fluctuations}, under the assumption that $\min_{k\neq k^*} \|\vec{\mu}_k-\vec{\mu}_{k^*}\| \rightarrow \infty$, the non-zero ordered sample eigenvalues $\ell_1{\>=} {\cdot}{\cdot}{\cdot}{\>=} \ell_{n-1}{>}0$ of matrix $\mat{S}_n$ can be separated into two sets by relating them to the eigenvalues of either $\mat{S}_\mathrm{B}$ or $\mat{S}_\mathrm{W}$ (graphically illustrated in Figure~\reffloat{fig:eg_bulk_spikes}{}):

\begin{enumerate}
	\item The major part of the non-zero eigenvalues $\ell_K{\>=} {\cdot}{\cdot}{\cdot}{\>=} \ell_{n-1}{>}0$ of $\mat{S}_n$, which are infinitely many as $n{\rightarrow}\infty$, will converge on a closed interval $[a,b] \subseteq (0,\infty)$ to the same limiting distribution of eigenvalues of the within-group covariance matrix $\mat{S}_\mathrm{W}$. This compact set of eigenvalues is therefore called the bulk.
	
	\item The top $K{-}1$ eigenvalues $\ell_1{\>=} {\cdot}{\cdot}{\cdot}{\>=} \ell_{K-1}$ of $\mat{S}_n$ will converge to certain limits $\lambda_1\>= {\cdot}{\cdot}{\cdot}{\>=} \lambda_{K-1} {>} b$, where $b$ is the upper bound of the bulk. The top $K{-}1$ eigenvalues are called the spikes which are induced by the $K{-}1$ eigenvalues of the between-group covariance matrix $\mat{S}_\mathrm{B}$.
\end{enumerate}

Note that the matrix $\mat{S}_p = \mat{M}\mat{M}^\T /p$ has the same non-zero eigenvalues as the matrix $\mat{S}_n$, but it is much easier to compute in practice as its dimension ($n\times n$) is much smaller than the dimension ($p \times p$) of $\mat{S}_n$. In what follows, we will use the matrix $\mat{S}_p$ to compute the sample eigenvalues, and we will also infer the number of sub-populations $K$ by performing spectral analysis on the matrix $\mat{S}_p$.	

As a consequence, for any finite number $m$ that satisfies $K\<=m\ll n$, the following results hold (almost surely) \citep{benaych2011eigenvalues},
\begin{equation*}
\begin{dcases}
\ell_i \rightarrow      \lambda_i,          & \aln{K}{1} \<= i \<= K-1, \\
\ell_i \rightarrow \aln{\lambda_i,{}}{b,{}} &      K     \<= i \<= m.
\end{dcases}
\end{equation*}
In particular, if we let $\lambda_K = b$ and define the ratio of the sample eigenvalue limit as $\theta_i = \lambda_{i+1}/\lambda_i$, then for the sample ERs $r_i = \ell_{i+1} / \ell_i$, we have the following convergence results

\begin{equation}
\begin{dcases}
r_i \rightarrow \aln{b/b}{\theta_i} < 1, & \aln{K}{1} \<= i \<= K-1, \\
r_i \rightarrow      b/b            = 1, &      K     \<= i \<= m. \\
\end{dcases}  
\label{eq:ER_limit}
\end{equation} 

To avoid confusion with bulk and spike, which are typically used for eigenvalues, we will refer to the ratio $r_i$ as the spiked ER when $1 \<= i \<= K{-}1$, and as the bulk ER when $K \<= i \<= n{-}1$ (the last ER $r_n$ is 0 by definition).

\subsection{The ER-based Estimator} 
\begin{figure}
	\centering
	\includegraphics[width=2.8in]{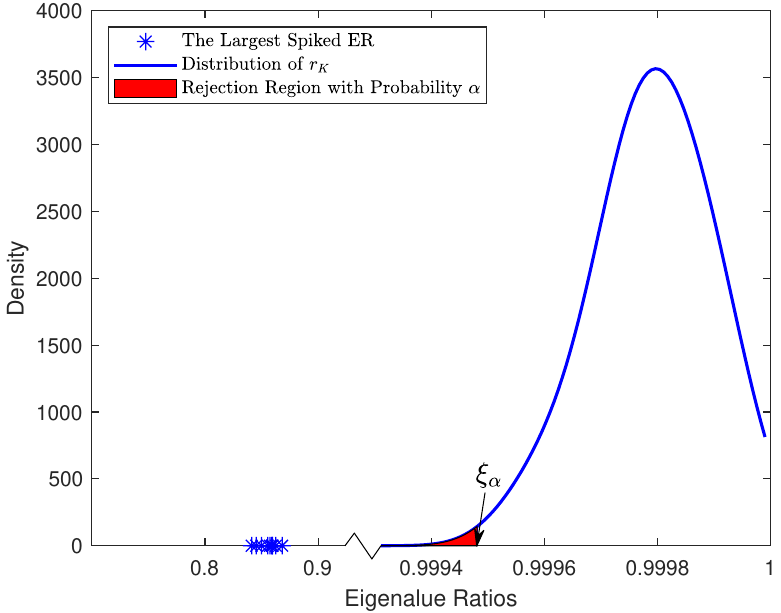}
	\caption{The largest spiked ER and the well-separated distribution of the top bulk ER $r_K$, generated from $100$ simulations following the same setting as in Figure~\reffloat{fig:eg_bulk_spikes}{}. Those ERs that are less than the critical value $\xi_\alpha$ will be considered as the spikes (here $\alpha$ is set as $0.005$ for illustration).}
	\label{fig:eg_FirstNoiseRatio}
\end{figure}

Based on the above asymptotic results, theoretically we can leverage the asymptotically consistent ER-based estimator studied in \citet{li2017identifying} to estimate the number of latent sub-populations $K$ as follows
\begin{equation}
\KER' \coloneqq \min\bigl\{ 1\<=i\<=n{-}1 \text{ \ s.t. \ } r_i > \xi_\alpha \bigr\}, \label{eq:K_ER_prime}
\end{equation}
where $\alpha$ is the pre-specified significance level, and the critical value $\xi_\alpha$ is chosen as the lower $\alpha$ quantile of the distribution of the top bulk ER $r_K$ (as illustrated in Figure~\reffloat{fig:eg_FirstNoiseRatio}{}). 

Note that by the definition in the Equation~\eqref{eq:K_ER_prime}{}, the probability of over-estimation (i.e., spurious detection on the bulk side) of our proposed estimator $\KER'$ is controlled by the significance level $\alpha$. However, there is no such control on the probability of under-estimation (i.e., spurious detection on the spiked side). Under-estimation might occur when the largest spiked ER $\max_{1\<=i<K} \{r_i\}$ jumps above the critical value $\xi_\alpha$, leading to an early stopping of our sequential testing procedure as given in Equation~\eqref{eq:K_ER_prime}{}. Thus, in order to control the probability of under-estimation, we need to know the theoretical joint distribution of the spiked ERs. But unlike the bulk, the asymptotic distribution of the spiked eigenvalues (and thus the spiked ERs) is sensitive and varies with different distributions of the entries in the data matrix \citep{benaych2011eigenvalues, benaych2011fluctuations}. As a result, each entry in the data matrix can substantially affect the distributions of the spikes. We observe that the last spiked ER should be asymptotically less than $\xi_\alpha$, while all the bulk ERs are greater than $\xi_\alpha$ with a probability controlled by $1{-}\alpha$ as in \eqref{eq:ER_limit}{}. With this observation, we can control the probability of under-estimation by stopping the sequential testing procedure only if all the $(\KER'+1)$th to $\hat{K}_c$th eigenvalue ratios are confirmed to be above the critical value. Here, $\hat{K}_c$ is some pre-specified coarse estimate $\hat{K}_c$ for the number of sub-populations, which should be generally larger than the true $K$. By default, we set $\hat{K}_c = n/10$ in our algorithm to ensure $\hat{K}_c>K$ because for usual statistical estimations (and especially in the ultra-dimensional scenarios), it is crucial to have at least $10$ samples in one sub-population to achieve decent estimation accuracy. 

Another challenge when applying our ER-based estimator defined in Equation~\eqref{eq:K_ER_prime}{} to real data sets is to find out a proper value of $\xi_\alpha$ as the distribution of $r_K$ is generally unknown. As we have mentioned in Section~\reffloat{sec:introduction}{}, the approach in \citet{li2017identifying} does not work on sequencing genotype data sets because of the severe LD and ultra-dimensionality issues. We now introduce a novel method to address these two issues and then obtain a more accurate approximation for $\xi_\alpha$.

According to~\citet{benaych2011fluctuations}, the distributions of the top two bulk eigenvalues $\ell_K$ and $\ell_{K+1}$ can be approximated by the distributions of the top two eigenvalues $\tilde\ell_1$ and $\tilde\ell_2$ of the following $n$-by-$n$ noise covariance matrix respectively:
\begin{equation}
\widetilde{\mat{S}}_p = \frac1p \mat{X}\mat{\Sigma}\mat{X}^\T, \label{eq:l2/l1}
\end{equation}
where $\mat{X}$ is a $n$-by-$p$ random matrix with all independent and identically distributed standard Gaussian entries. Let $\tilde{r}_1 = \tilde\ell_2/\tilde\ell_1$, the above result implies that $r_K \distApprox \tilde{r}_1$, where the notation $\distApprox$ denotes that the two random variables on the two sides asymptotically follow the same distribution. Given the population covariance matrix $\mat{\Sigma}$ and a significance level $\alpha$, we can in principle approximate $\xi_\alpha$ using the equation $\mathrm{P}(0 < \tilde{r}_1 \<= \tilde\xi_\alpha) = \alpha$.

However, even if we know the true covariance $\mat{\Sigma}$, the top two eigenvalues $\tilde\ell_1$ and $\tilde\ell_2$ are essentially roots of a polynomial equation of order $n$ whose distribution functions have no closed-form expressions in general. It is also computationally inefficient to use Monte Carlo method to simulate the null distribution based on the covariance matrix $\widetilde{\mat{S}}_p$ defined in Equation~\eqref{eq:l2/l1}{} as we need to generate a huge $n$-by-$p$ matrix $\mat{X}$ multiple times, where $p$ is at the order of millions. To solve this problem, we apply the limiting theory for the eigenvalues of the sample noise matrix developed by~\citet{wang2014limiting} under the Asymptotic Regime~\hyperref[regime]{\bfseries A}. The theory states that the eigenvalues of $\sqrt{p/nb_p} \bigl( \widetilde{\mat{S}}_p - a_p\mat{I}_n \bigr)$ converge almost surely to the \emph{semicircle law}, where $a_p = \mathrm{tr}\bigr( \mat{\Sigma} \bigl)/p$ and $b_p = \mathrm{tr}\bigr( \mat{\Sigma}^2 \bigl)/p$ and $\mathrm{tr}(\cdot)$ denotes the trace of a matrix. The semicircle law gives the same limiting distribution for the eigenvalues of $\mat{W}_n/\sqrt{n}$ when $n\rightarrow\infty$, where $\mat{W}_n$ is a $n$-by-$n$ Gaussian orthogonal ensemble (GOE) matrix (i.e., a square matrix with independent entries where each diagonal entry follows $N(0,2)$ and each off-diagonal entry follows $N(0,1)$) \citep{wigner1958distribution, arnold1971wigner}. Hence, the relationships of the top two eigenvalues of $\sqrt{p/n b_p}\bigl( \widetilde{\mat{S}}_p - a_p\mat{I}_n \bigr)$ and $\mat{W}_n/\sqrt{n}$ are given by 
\begin{equation}
\sqrt{p/nb_p} \cdot (\tilde\ell_i - a_p) \distApprox w_i/\sqrt{n}, \qquad i=1,2,	\label{eq:eigen_approx}
\end{equation}
where $w_1$ and $w_2$ are the top two eigenvalues of the matrix $\mat{W}_n$. As a result, the top bulk ER $r_K$ can be approximated by
\begin{gather}
	r_K \distApprox
	\tilde{r}_1 = \dfrac{\tilde\ell_2}{\vphantom{\sqrt{\hat{b}_p}}\tilde\ell_1} \distApprox
	\dfrac{w_2 {\cdot} \sqrt{\hat{b}_p/p} + \hat{a}_p}{\aln{w_1}{w_1} {\cdot} \sqrt{\hat{b}_p/p} + \hat{a}_p} = \tilde{r}_1^*, 
	\label{eq:ER_approx}
\end{gather}
where
\begin{gather}
	\hat{a}_p = \frac1{n-K} \sum_{i=K}^{n-1} \ell_i, \qquad
	\hat{b}_p = \frac{\aln{n^2}{p}}{(n-K)^2} \sum_{i=K}^{n-1} (\ell_i - \hat{a}_p)^2, \label{eq:ap_bp}
\end{gather}

are the two moment estimators for $a_p$ and $b_p$ in Equation~\eqref{eq:eigen_approx}{} respectively.

Denote $\tilde\xi_\alpha^*$ as the new approximation for the critical value $\xi_\alpha$ in Equation~\eqref{eq:K_ER_prime}{}. Given a pre-specified coarse estimator $\hat{K}_c<n{-}1$, our final ER-based estimator for the number of sub-populations $K$ is
\[
	\KER \coloneqq \min\bigl\{ 1\<=i\<=\hat{K}_c \text{ \,s.t.\, } r_i, \dotsc, r_{\!{\scriptscriptstyle\hat{K}\!}_c} \text{ all greater than } \tilde\xi_\alpha^* \bigr\}, 
\]
where we choose $\tilde\xi_\alpha^*$ as the lower $\alpha$ quantile of the distribution of $\tilde{r}_1^*$ in Equation~\eqref{eq:ER_approx}{}, that is,
\[
	\mathrm{P}( 0 < \tilde{r}_1^* \<= \tilde\xi_\alpha^* ) = \alpha.
\]
\vspace{-\baselineskip}

\subsection{ERStruct Algorithm}
We summarize our method as an algorithm to estimate $K$, the number of top informative PCs that capture the latent population structure, from a raw genotype data matrix $\mat{C}$ below.

\begin{breakablealgorithm}
	\caption{ERStruct}
	\begin{algorithmic}[1] 
		\Require $\begin{aligned}[t]
			\aln{\hat{K}_c}{\mat{C}}:\quad & n\times p \text{ genotype data matrix} \\[-\smallskipamount]
			     \hat{K}_c:          \quad &\text{a coarse estimate (set to } \lfloor n{/}10 \rfloor \text{ by default)} \\[-\smallskipamount]
			\aln{\hat{K}_c}{m}:      \quad &\text{the number of Monte Carlo replicates} \\[-\smallskipamount]
			\aln{\hat{K}_c}{\alpha}: \quad &\text{significance level} \\[-\smallskipamount]
		\end{aligned}$ 
		\Ensure $\KER$:\quad ER estimation of the number of top informative PCs
		
		\State $\mat{M} \gets$ Equation~\eqref{eq:normalization}{}
		\State $\mat{S}_p \gets \frac1p\mat{M}\mat{M}^\T$
		\State $\ell_1{\>=}\ell_2{\>=}{\dotsb}{\>=}\ell_{n-1} \gets$ ordered non-zero eigenvalues of $\mat{S}_p$
		\State $(r_1,\dotsc,r_{n-2}) \gets (\ell_{2}/\ell_1, \dotsc, \ell_{n-1}/\ell_{n-2})$
		
		\State $\bigr((w_1^{(1)} \cdots w_1^{(m)})^\T, (w_2^{(1)} \cdots w_2^{(m)})^\T\bigl)$ $\gets$ generate $m$ replicates of the top two eigenvalues of GOE matrices $\mat{W}_n$
		
		\For{$K \gets 1$ \textbf{to} $\hat{K}_c$}
			\If{no valid $\KER$}
				\State $(\hat{a}_p,\ \hat{b}_p) \gets$ Equations~\eqref{eq:ap_bp}{}
				\For{$i \gets 1$ \textbf{to} $m$}
					\State $(w_1, w_2) \gets (w_1^{(i)}, w_2^{(i)})$
					\State $\tilde{r}_1^{*(i)} \gets$ Equation~\eqref{eq:ER_approx}{}
				\EndFor
				\State $(\prescript{\uparrow}{}{\tilde{r}}_1^{*(1)}, \dotsc, \prescript{\uparrow}{}{\tilde{r}}_1^{*(m)}) \gets$ sort $(\tilde{r}_1^{*(1)}, \dotsc, \tilde{r}_1^{*(m)})$ in ascending order
				\State $\tilde\xi_\alpha^* \gets \prescript{\uparrow}{}{\tilde{r}}_1^{*(\lceil m\alpha \rceil)}$
				\If{$r_K > \tilde\xi_\alpha^*$}
					\State $\KER \gets K$ and set $\KER$ as valid
				\EndIf
			\ElsIf{$r_K \<= \tilde\xi_\alpha^*$} set $\KER$ as not valid
			\EndIf
		\EndFor
	\end{algorithmic}
\end{breakablealgorithm}

\section{Simulation Studies} \label{sec:simulation_studies}
In this section, we compare the performances of our new method ERStruct versus the original TW test \citep[i.e., PCA-TW test without using the estimated effective number of markers; see][]{patterson2006population} and the PCA-TW test by \citet{patterson2006population}. The comparison is made based on how close their estimated number of PCs are to the ground truth. We consider the following two different settings.

\begin{enumerate}
	\item In the uncorrelated (no LD) setting, the marker-to-marker covariance is set as $\mat{\Sigma} = 0.5{\cdot}\mat{I}_p$, and $100$ independent Monte Carlo replicated samples of genotype data matrices are generated according to the ANOVA model \eqref{eq:ANOVA}{}. The other parameters are set so that the simulated data is similar to the 1000 Genomes Project data set with MAF less than $5\%$ genetic markers filtered out as analyzed in Section~\reffloat{sec:real_data_analysis}{}. Specifically, the number of markers $p=7{,}921{,}816$; the number of individuals $n=2504$; the number of sub-populations $K=26$; the numbers of individuals in each sub-population $n_k=\{\myvec{96, 61, 86, 93, 99, 103, 105, 94, 99, 99, 91, 103, 113, 107, 102, 104, 99, 99, 85, 64, 85, 96, 104, 102, 107, 108}\}$; the noise vector $\vec{\varepsilon}_{k,l} \sim N(\vec{0},0.5{\cdot}\mat{I}_p)$; and the mean count of minor alleles $\vec{\mu}_k \sim \mathrm{Binomial}(2,\hat{\vec{\mu}}_k/2)$, with $\hat{\vec\mu}_k$ obtained by Equation~\eqref{eq:mu_hat}{} in which $\vec{c}_{k,l}$ is the raw minor allele counts of the $l$th individual in the $k$th sub-population from the 1000 Genomes Project data set. Finally, we use a rounding mapping $x \mapsto I_{x\>=1.5}(x) - I_{x<0.5}(x) + 1$ so that all the simulated genotype data take values in $\{0,1,2\}$, where $I_A(\cdot)$ denotes an indicator function on a set $A$.
	
	\item In the LD setting, in order to simulate local marker-to-marker correlations (the LD matrix), the noise vectors $\vec{\varepsilon}_{k,l}$ within the $k$th sub-population are generated from the distribution $N(\vec{0},0.5{\cdot}\mat{\Sigma}_k)$. Each $\mat{\Sigma}_k$ is a block diagonal matrix extracted from the sample correlation matrix of the $k$th sub-population in the 1000 Genomes Project data set. All the other parameters (i.e., the number of markers $p$, the number of individuals $n$, the number of sub-populations $K$, the numbers of individuals in each sub-population $n_k$ and the mean count of minor alleles $\vec{\mu}_k$) are set to be the same values as in the uncorrelated setting.
\end{enumerate}

\begin{figure}
	\centering
	\includegraphics[width=0.95\textwidth]{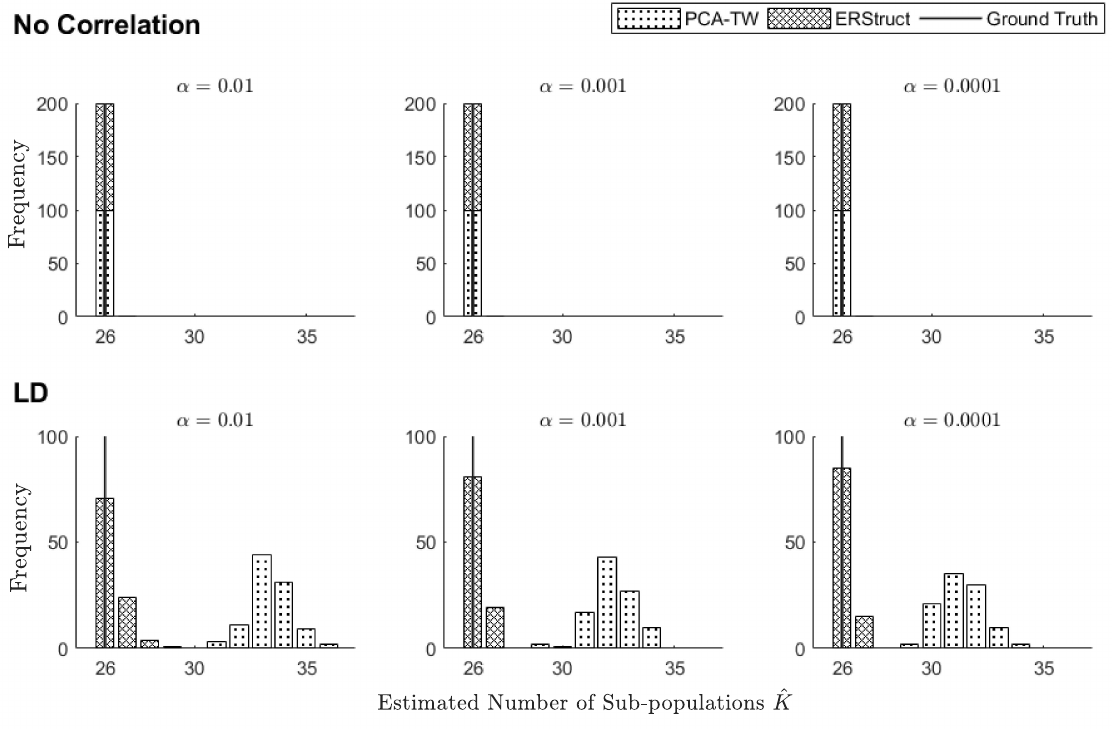}
	\caption{Histograms of the estimated number of sub-populations using the PCA-TW test (dotted) and ERStruct (crossed) in uncorrelated (upper panel) and LD settings (lower panel), each with 100 replications. The ground truth is $K=26$.}
	\label{fig:simulation}	
\end{figure}

\begin{table}
	\caption{Simulation results for the estimated numbers of sub-populations using the PCA-TW test and ERStruct in the LD setting. The correct estimations represent the percentages of estimations that are equal to the true number (26) of sub-populations. The range denotes the minimum and maximum estimated numbers.}
	\centering
	\begin{tabular*}{\textwidth}{l@{\extracolsep{\fill}}c@{\extracolsep{\fill}}c@{\extracolsep{\fill}}c@{\extracolsep{\fill}}c@{\extracolsep{\fill}}c@{\extracolsep{\fill}}c}
		\toprule
		& \multicolumn{3}{c}{\bfseries{PCA-TW test}} & \multicolumn{3}{c}{\bfseries{ERStruct}} \\
		\cmidrule{2-4} \cmidrule(r){5-7}
		$\alpha$ level & 0.01      &   0.001   &          0.0001          & 0.01      &   0.001   &        0.0001         \\
		\midrule
		correct estimations                               & $0$       &    $0$    &           $0$            & $71\%$    &  $81\%$   &        $85\%$         \\
		range                                             & $[31,36]$ & $[29,34]$ &        $[29,34]$         & $[26,29]$ & $[26,27]$ &       $[26,27]$       \\
		bias                                              & $7.38$    &  $6.22$   &          $5.31$          & $0.35$    &  $0.19$   &        $0.15$         \\
		variance                                          & $0.94$    &  $1.02$   &          $1.08$          & $0.37$    &  $0.16$   &        $0.13$         \\
		\bottomrule
	\end{tabular*}
	\label{tab:summary_stats}
\end{table}

The results of using the original TW-test are far from the ground truth number ($K=26$), as expected. With the smallest significance level $\alpha=0.0001$ we considered, the numbers of top PCs found by the original TW test are in the range $(1918,1921)$ among the 100 Monte Carlo replicates in the uncorrelated setting, and it gets even worse in the LD setting where the numbers of top PCs are in the range of $(2057,2081)$. These simulation results show  that both ultra-dimensionality and LD should be taken into account  in order to accurately capture population structure in sequencing data sets.

The simulation results of using the PCA-TW test and our ERStruct method are shown in Table~\reffloat{tab:summary_stats}{} and Figure~\reffloat{fig:simulation}{}. Our proposed ERStruct method outperforms the PCA-TW test.  Specifically, in the uncorrelated setting which favors the PCA-TW test, our ERStruct achieve the same accuracy as the PCA-TW test. In the LD setting, 85\% ($\alpha=0.0001$) of the replicates using ERStruct are correctly estimated, and the remaining 15\% of the replicates still give highly accurate estimates (ground truth $K=26$ versus $\KER=27$ when $\alpha=0.0001$). In contrast, the PCA-TW test has less accurate estimates among those 100 replicates and the estimates are far away from the ground truth (ranging from $29$ to $34$ even when $\alpha=0.0001$). Our ERStruct method still performs well even if the covariance matrix $\mat{\Sigma}_k$ varies across sub-populations.

It is worth noting that the ERStruct method is developed under the assumption that the populations share a common covariance matrix $\mat\Sigma$ while in the simulation setting with LD, we used different covariance matrices across the sub-populations. The results show that the ERStruct method is robust against this variability. One possible explanation to this robustness is that the method relies on two estimates for population spectral moments $a_p$ and $b_p$ given in Equation~\eqref{eq:ap_bp}{}. It is very likely that these two estimates, derived in the case of a constant covariance matrix across all sub-populations, are still accurate for the case of different covariance matrices. However, giving a rigorous justification to this claim is difficult and out of reach from the current state of random matrix theory.

\section{Real Data Analysis} \label{sec:real_data_analysis}
In this section, we apply our proposed ERStruct to two publicly available genotype data sets to estimate the number of top informative PCs for illustration. We also include the PCA-TW method for comparison purposes.

The first data set is from the HapMap 3 project \citep{international2010integrating}, which is a large-scale array-based genotype data set that includes $1115$ individuals from $11$  sub-populations around the world (see Web Table~S1 for more detailed geographical information). Although the primary interest of this paper is focused on large-scale whole genome sequencing data, HapMap 3 includes  a large number of markers ($1{,}615{,}203$ for the raw data in total) such that  the sample-to-marker ratio of the raw data is sufficiently small ($n/p = 6.9 {\times} 10^{-4}$). Therefore, the HapMap 3 data set can be considered as falling into our ultra-dimensional Asymptotic Regime~\hyperref[regime]{\bfseries A}, and can serve as a good example to illustrate our method. 

The second data set is from the 1000 Genomes Project \citep{10002015global}, which is a whole-genome sequencing data set with  $2504$ individuals from $26$ sub-populations (detailed geographical information is given in Web Table~S2). It was established in January 2008 with the aim to build by then the most detailed catalog of genetic variations in the human population. The 1000 Genomes Project inherits the major part of the data in HapMap 3, with additional featured sub-populations and a lot of rarer genetic variants ($81{,}271{,}745$ for the raw data in total) included.

Our raw data pre-processing is as follows. We first filter genetic markers using the PLINK software \citep{purcell2007plink} by imposing different levels of MAF filtering thresholds. For HapMap 3, markers with MAF less than ($5\%$, $1\%$) are removed. For the 1000 Genomes Project sequencing data set, we also investigate several situations in which we remove genetic markers with MAF less than ($5\%$, $1\%$, $0.5\%$, $0.1\%$, $0.05\%$, $0.01\%$). To further investigate the performance of our ERStruct method on LD-pruned data, we performed LD-pruning on the 1000 Genomes Project data  by removing genetic markers with MAF less than ($5\%$, $1\%$) and LD $r^2 >0.1$ in a window size of 10,000 base pair (bp). Then under multiple significance $\alpha$ levels ($0.01$, $0.001$, $0.0001$), we applied the traditional PCA-TW method and our ERStruct method to the pre-processed data sets. 

\begin{figure}
	\centering
	\includegraphics[width=.75\textwidth]{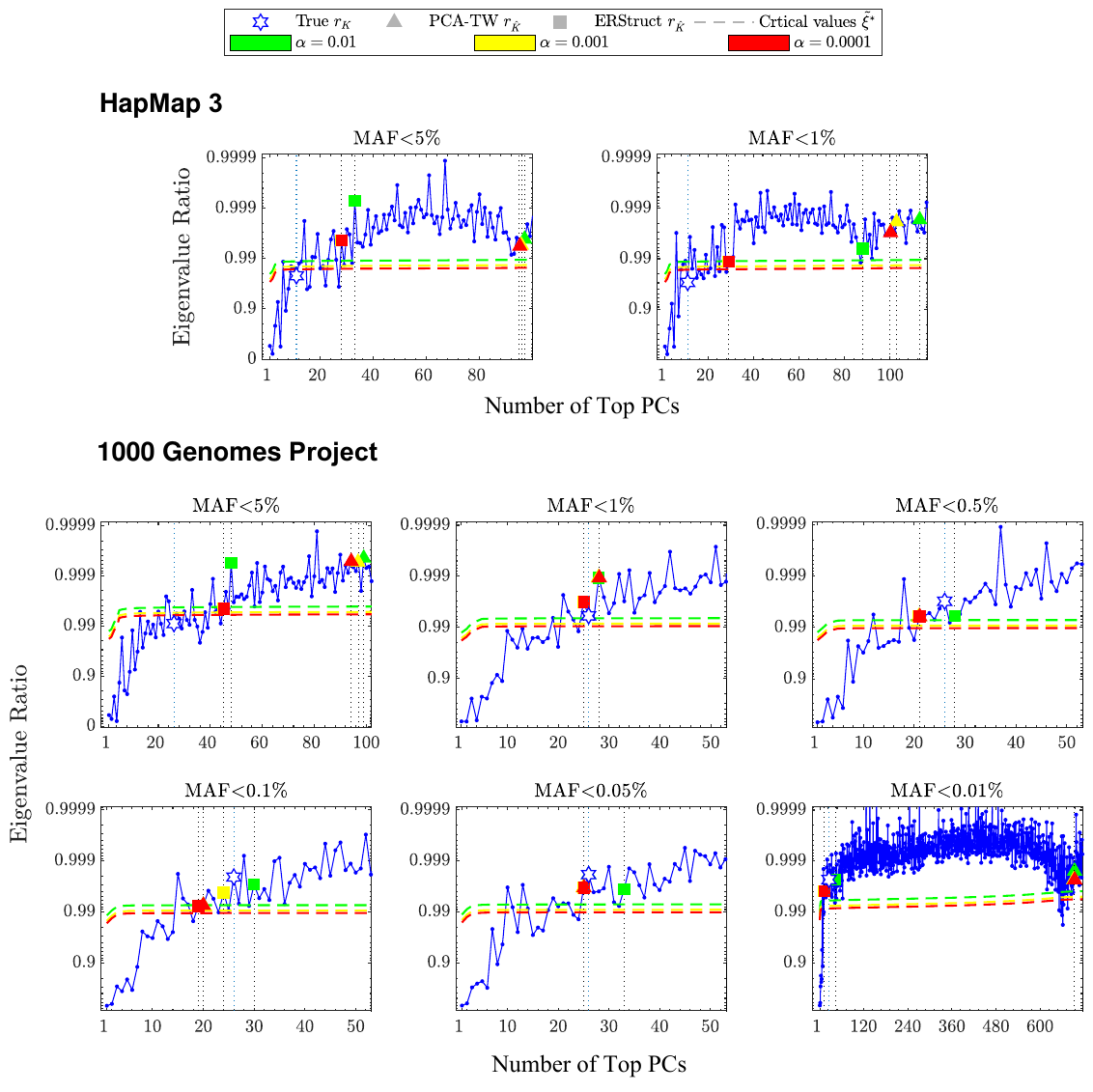}
	\caption{Scree plots of ERs computed from the HapMap 3 ($11$ sub-populations) and 1000 Genomes Project ($26$ sub-populations) data sets. Each sub-plot is created based on different MAF filtering thresholds on the raw data, in which the x-axis represents the number of top PCs, and the y-axis represents the corresponding $x$th top ER with a log-scale transformation $y \mapsto -\log_{\scriptscriptstyle \hspace{-.5pt}1\hspace{-.5pt}0\hspace{-.5pt}}(1{-}y)$ to increase visibility. True bulk ERs that correspond to the real number of sub-populations in each data set are shown in the shape of hexagram, while top bulk ERs that correspond to estimations with significance levels $\alpha=0.01$ (red), $0.001$ (yellow) and $0.0001$ (green) are emphasized in the shape of triangle (PCA-TW test) and square (ERStruct). This figure appears in color in the electronic version of this article, and any mention of color refers to that version.}
	\label{fig:real_data_eigratio}
\end{figure}

\begin{table}
	\caption{Comparison of estimations based on different filtering thresholds of the raw HapMap 3 (1{,}115 individuals), the raw 1000 Genomes Project (2{,}504 individuals), and the LD-pruned 1000 Genomes Project ($r^2<0.1$ in a window size 10,000 bp) data sets using the PCA-TW test and ERStruct method with significance levels $\alpha=0.01$, $0.001$ and $0.0001$.}
	\centering\small
	\begin{tabular}{l@{\qquad}ccc@{\qquad}
			>{\centering}p{\widthof{0.01}}@{\hspace{0.7em}}
			>{\centering}p{\widthof{0.001}}@{\hspace{0.7em}}
			>{\centering\hspace{-.5em}}p{\widthof{0.0001}}
			@{}
			>{\centering}p{\widthof{0.01}}@{\hspace{0.7em}}
			>{\centering}p{\widthof{0.001}}@{\hspace{0.7em}}
			>{\centering\hspace{-.5em}\arraybackslash}p{\widthof{0.0001}}
		}
		\toprule
		& & & & \multicolumn{3}{c}{\bfseries{PCA-TW test}\hspace*{1em}} & \multicolumn{3}{c}{\bfseries{ERStruct}\hspace*{1em}} \\
		\cmidrule(r{1.5em}){5-7} \cmidrule(r){8-10}
		& MAF filter & no. of markers & $n$-to-$p$ ratio & 0.01 & 0.001 & 0.0001 & 0.01 & 0.001 & 0.0001  \\
		\midrule
		\multirow[t]{2}{*}{HapMap 3}     & $     \text{MAF}<5\%                             $ & \phantom01{\,}493{\,}644 &  $7.5\times10^{-4}$ & $\phantom097$ &  $\phantom096$ &  $\phantom095$ & $33$ & $28$ & $28$ \\
		& $     \text{MAF}<1\%                             $ & \phantom01{\,}601{\,}085 & $7.0\times10^{-4}$ & $113$ & $103$ & $100$ & $88$ & $29$ & $29$ \\[\smallskipamount]
		\multirow[t]{2}{*}{\parbox[t]{\widthof{1000 Genomes}}{1000 Genomes Project \rlap{(pruned)}}}    & $     \text{MAF}<5\%                    $ & \phantom{00\,}161{\,}842 &  $1.5\times10^{-2}$ & $\phantom061$ &  $\phantom060$ &  $\phantom059$ & $52$ & $52$ & $46$ \\
		& $     \text{MAF}<1\%                 $ & \phantom{00\,}388{\,}636 & $6.4\times10^{-3}$ & $116$ & $110$ & $107$ & $57$ & $53$ & $53$ \\[\smallskipamount]
		\multirow[t]{4}{*}{\parbox[t]{\widthof{1000 Genomes}}{1000 Genomes Project}} & $     \text{MAF}<5\%                             $ & \phantom07{\,}921{\,}816 & $3.2\times10^{-4}$ & $\phantom099$ &  $\phantom097$ &  $\phantom094$ & $48$ & $45$ & $45$ \\
		& $     \text{MAF}<1\%                             $ & 13{\,}650{\,}478 & $1.8\times10^{-4}$ & $\phantom028$ &  $\phantom028$ &  $\phantom028$ & $28$ & $25$ & $25$ \\
		& $     \text{MAF}<0.5\%                           $ & 17{\,}307{\,}567 & $1.4\times10^{-4}$ & $\phantom021$ &  $\phantom021$ &  $\phantom021$ & $28$ & $21$ & $21$ \\
		& $     \text{MAF}<0.1\%                           $ & 28{\,}793{\,}505 & $8.7\times10^{-5}$ & $\phantom020$ &  $\phantom020$ &  $\phantom020$ & $24$ & $24$ & $19$ \\
		& $     \text{MAF}<0.05\%                          $ & 37{\,}961{\,}945 & $6.6\times10^{-5}$ & $\phantom026$ &  $\phantom025$ &  $\phantom025$ & $33$ & $25$ & $25$ \\
		& $     \text{MAF}<0.01\%                          $ & 81{\,}017{\,}519 & $3.1\times10^{-5}$ & $694$ & $693$ & $693$ & $43$ & $13$ & $13$ \\
		\bottomrule 
	\end{tabular}
	\label{tab:Hampmap_1KGP}
\end{table}

Since both Hapmap 3 and 1000 Genomes projects consist of human samples from known sub-populations, the number of sub-populations in each of the two data sets can be regarded as the ground truth. The estimated number of top informative PCs is expected to be close to the ground truth. As shown in Figure~\reffloat{fig:real_data_eigratio}{}, all the ER scree plots  oscillate with no specific patterns around the true $r_K$. This shows that the simple eyeballing method is not accurate here, and we need to estimate the number of top informative PCs  using statistically more rigorous  methods. More detailed estimation results are summarized in Table~\reffloat{tab:Hampmap_1KGP}{}, including results on the LD-pruned 1000 Genomes Project data. Under  all settings, our ERStruct gives very accurate estimates. In comparison, the PCA-TW test gives a severe over-estimation on the 1000 Genomes Project data set when the MAF filtering thresholds are  $5\%$ and  $0.01\%$. The problem of over-estimation remains in all filtering thresholds of the smaller HapMap 3 data set, which has sample-to-marker ratios closer to the asymptotic regime assumed in the PCA-TW test ($n/p \rightarrow constant >0$). These results are also in line with our findings in Section~\reffloat{sec:simulation_studies}{}, and show that our ERStruct method is more robust to the present of LD and different MAF filtering thresholds under consideration. Even if there is a certain degree of information loss after LD pruning, our ERStruct method still performs better than the PCA-TW method.

To further assess the empirical performance of our ERStruct method, we adopt the following cross-validation procedure. We first randomly sampled $30\%$ individuals from each sub-population in the original data matrix as the testing data, and the remaining $70\%$ as the training data. We obtain the normalized data $\mat{M}_\text{train}$ and $\mat{M}_\text{test}$ through Equation~\eqref{eq:normalization}{}, respectively. Then we choose the first $k$ PCA loadings $\mat{V}_k$ computed from the normalized training data $\mat{M}_\text{train}$ and then we try to recover the normalized testing data by $\widetilde{\mat{M}}_k = \mat{V}_k \mat{V}_k^\T \mat{M}_\text{test}$. The recovered testing data $\widetilde{\mat{M}}_k$ should be close to the original testing data $\mat{M}_\text{test}$ if the top $k$ selected PCs are sufficiently informative for capturing population structure. 

We used the metric $\delta_k = \Vert \widetilde{\mat{M}}_k - \mat{M}_\text{test} \Vert_1$ to measure how close the recovered testing data matrix is to the original testing data matrix, where $\Vert\cdot\Vert_1$ is the induced matrix 1-norm defined as $\Vert\mat{A}\Vert_1 \coloneqq \max_{1\<=j\<=p} \sum_{i=1}^p |a_{ij}|$. As an example, we plot the metric $\delta_k$ versus the number of selected top PCs $k$ using the 1000 Genomes Project data with $\text{MAF}<5\%$ markers removed (see Web Figure~S1). We found that even though the overall trend of the curve is decreasing as $k\rightarrow n$, there is clearly a local ``valley'' in the range $(20,50)$, suggesting that a good choice of top informative PCs should be in this range. We can also see from the curve that the estimated number of  top informative PCs using our ERStruct method fell into this range  when the significance levels are $\alpha=(0.01,0.001,0.0001)$, and the metric  $\delta_k$ is smaller (better recovery of the testing data matrix) in comparison with the PCA-TW test.

The above real data analysis results show that our ERStruct method is more accurate and robust compared to the PCA-TW test. Based on our observations, we recommend a filtering threshold for removing genetic markers with MAF less than $1\%$ and a significance level of 0.001, as the default parameters setting when applying our ERStruct method on whole genome sequencing data sets. In addition to this empirical recommendation, users are also suggested to perform sensitivity analysis by varying the MAF filtering threshold and significance levels.

\section{Discussion} \label{sec:discussion}
In this paper, we proposed a new method ERStruct to estimate the number of top informative PCs in whole genome sequencing data accounting for complicated LD structure between genetic markers. Our ERStruct method has been shown to outperform the traditional PCA-TW test in both simulated and real  data sets. This demonstrates that our ERStruct method has wide applications to the increasingly available whole genome sequencing data sets to infer population structure \citep{bycroft2018uk,wu2019large,cao2020chinamap}.

Our ERStruct method enjoys several advantages. First, our ERStruct estimator is based on the more robust eigenvalue ratios when LD is present. Second, we obtain a more accurate adaptive null distribution approximation for the ER test statistic under the ultra-dimensional regime which is specifically developed for modern sequencing data. Third, our method is not confined to a specific LD structure among genetic markers. Even though in the real genetic data with complicated LD structures, our ERStruct can still separate the ER spikes from the ER bulk. Fourth, our ERStruct method is also computationally efficient. In fact, our ERStruct achieves almost the same computational speed as the PCA-TW test, given that $p \gg n$. For example, it took only around 30 minutes to obtain the estimate of the 1000 Genomes Project data set with MAF less than $0.05\%$ removed ($37{,}961{,}945$ markers) using our ERStruct MATLAB toolbox on a server with 126G RAM and 5 cores of CPU.

Our proposed ERStruct method can also be extended to infer latent structures (like the number of latent batches) in other types of ultra-dimensional genomic data. For example, in single-cell sequencing data, most of the entries in the data matrix are zeros. Inference of latent structures in such zero-inflated sparse data matrix is still very challenging because  the null distribution of the ER test statistic might be distorted (\citealp*{jong2019local}; \citealp{aparicio2020aRandom}). More future work is needed to  extend our ERStruct method for such zero-inflated data matrices.


\backmatter


\section*{Acknowledgements}
We thank the editor, associate editor and reviewer for their valuable comments which improved this paper. 
Dr. Zhonghua Liu is supported by Hong Kong Research Grants Council Early Career Scheme (27307920).
\vspace*{-8pt}

\section*{Data Availability}
The data sets used in this paper are openly available at the following links, \\[\smallskipamount]
\citet{international2010integrating}: \\
\indent\refurl{https://www.sanger.ac.uk/resources/downloads/human/hapmap3.html}. \\[\smallskipamount]
\citet{10002015global}: \\
\indent\refurl{https://www.internationalgenome.org/data}.
\vspace*{-8pt}



\bibliographystyle{biom} 
\bibliography{ERStruct}

%
%
%


\section*{Supporting Information}
Web Appendices, Tables, and Figures referenced in Section~\reffloat{sec:real_data_analysis}{} are available with this paper at the Biometrics website on Wiley Online Library. A MATLAB toolbox implementing our ERStruct algorithm is available at \refurl{https://github.com/bglvly/ERStruct} including code and example data. The source code is also available at the Biometrics website on Wiley Online Library.
\vspace*{-8pt}

%
%
%

\label{lastpage}

\end{document}